\documentstyle[12pt]{article}
\evensidemargin \oddsidemargin
\def\beq{\begin{equation}}
\def\eeq{\end{equation}}
\def\bea{\begin{eqnarray}}
\def\eea{\end{eqnarray}}
\def\ba{\begin{array}}
\def\ea{\end{array}}
\def\del{\partial}
\def\b{\bar}
\def\t{\widetilde}
\def\d{\dagger}

\def\g{\gamma}

\def\al{\alpha}

\begin{document}
\pagestyle{empty}
\begin{flushright}
CERN-TH/97-32\\
IC/97-23\\
hep-th/9703163\\
\end{flushright}
\vspace*{1cm}
\begin{center}
\vspace{.5cm}
{\bf \LARGE D-brane Black Holes: Large-N Limit and the} \\
\vspace{.5cm}
{\bf\LARGE Effective String Description} \\
\vspace*{1cm}
{\bf S. F. Hassan}{\footnote{\tt e-mail: fawad@ictp.trieste.it}} \\
\vspace{.2cm}
High Energy Physics Group, ICTP, \\
34100 Trieste, Italy\\
\vspace{.5cm}
{\bf Spenta R. Wadia}{\footnote{\tt e-mail: wadia@nxth04.cern.ch, 
wadia@theory.tifr.res.in}}{\footnote{On leave from the Tata  
Institute of Fundamental Research, Homi Bhabha Road, Mumbai-400005,
India}} \\    
\vspace{.2cm}
Theoretical Physics Division, CERN \\
CH - 1211 Geneva 23, Switzerland \\
\vspace*{1cm}
{\bf Abstract}
\end{center}
\begin{quote}
We address the derivation of the effective conformal field theory
description of the 5-dimensional black hole, modelled by a collection
of $D1$- and $D5$- branes, from the corresponding low energy
$U(Q_1)\times U(Q_5)$ gauge theory.  Finite horizon size at weak
coupling requires both $Q_1$ and $Q_5$ to be large. We derive the
result in the moduli space approximation (say for $Q_1>Q_5$) and
appeal to supersymmetry to argue its validity beyond weak coupling. As
a result of a combination of quenched $Z_{Q_1}$ Wilson lines and a 
residual Weyl symmetry, the low-lying excitations of the 
$U(Q_1)\times U(Q_5)$
gauge theory are described by an effective N=4 superconformal field
theory with $c=6$ in 1+1 dimensions, where the space is a circle of
radius $RQ_1Q_5$. We also discuss the appearance of a marginal
perturbation of the effective conformal field theory for large but
finite values of $Q_5$.
\end{quote}
\vfill
\begin{flushleft}
CERN-TH/97-32\\
IC/97-23\\
March 1997\\
\end{flushleft}
\vfill\eject
\pagestyle{plain}
\section{Introduction}
The identification of D-branes as RR charged solitons in \cite{POL} has
added new substance to string theory. One important physical
application is to black hole physics. Various black hole solutions of
low energy string theory admit a constituent description in terms of
D-branes \cite{STROMVAFA}. For extremal and near extremal black holes
the constituent model gives a statistical basis to the notion of black
hole entropy and leads to an understanding of the Hawking-Bekenstein
formula \cite{STROMVAFA,HORSTROM,JKM,BRECKEN,CALMALDA,MALDASUSS}.  In
\cite{STROMVAFA} the idea of the effective conformal field theory that
accounted for the entropy of the black hole was introduced. In
\cite{CALMALDA} the applications of the D-brane model to black hole
thermodynamics were initiated and in \cite{MALDASUSS} the
thermodynamics was explained in terms of the effective conformal field
theory that lives on a cylinder whose radius is dilated by a factor
related to the charges of the black hole. The formulation of
\cite{MALDASUSS} was used in \cite{DMW} to show that the low frequency
absorption rates for minimally coupled scalars calculated using
D-branes were proportional to the classical calculation. This work
also established that the D-brane assembly in the classical limit was
indeed ``black''. At that time it was not clear whether these rates
should be equal because the D-brane calculations were done in a regime
of couplings where the event horizon would fall within the string
length.  Subsequently the equality of these rates was shown in
\cite{DASMATH} in the weak coupling regime. The absorption
calculations were extended to the highly non-trivial case of grey body
factors in \cite{MALDASTROM} and also other circumstances in
\cite{CALGUBKLEB,KLEBKRAS,KLEBMATH}. There are recent works that also
discuss disagreements of the D-brane calculations with the classical
results \cite{JENNIE,ARVIND} and these issues are still to be
understood.

It is clear that the remarkable agreements between weak coupling
D-brane calculations and the classical results for near extremal black
holes need to be explained in terms of the standard theory of the
D-brane system \cite{WITTEN, POL} which is in principle valid for both
weak and strong coupling, when the branes are very near each
other. There is also the suggestion of a supersymmetry
non-renormalization theorem at work \cite{MALDASTROM} and an attempt
to derive, on general grounds, the low energy effective sigma model
that describes the collective modes of the black hole
\cite{MALDASTROM}.  The issue of string loop corrections to the
Dirac-Born-Infeld action has also been recently addressed \cite{das}.

In this work we attempt to address, in a more systematic way, the
issue of the description of the 5-dimensional black hole arising in
toroidal compactification of type IIB string theory in terms of an
assembly of D-branes. The black hole we consider is the one
considered in this connection in \cite{CALMALDA} and carries charges
$Q_1$, $Q_5$ and $n$. The D-brane system corresponding to this black
hole is described by an N=4 supersymmetric $U(Q_1)\times U(Q_5)$ gauge
theory in $1+1$ dimension with the space dimension as a circle of
radius $R$. This theory is the dimensional reduction of an N=2 theory
in $3+1$ dimensions \cite{MALDATHESIS,HASHI}.  The low energy
collective modes of the gauge theory have an effective description in
terms of a superconformal field theory in 1+1 dimensions. The black
hole is then identified with a certain state in this conformal field
theory and the degeneracy of this state determines the black hole
entropy. In this paper, we analyze the supersymmetric gauge theory
corresponding to the black hole and obtain the effective theory for
its low-lying excitations in an explicit way.

This paper is organized as follows: In section 2, using some
observations made in \cite{MALDASTROM} we recognise, in analogy with
QCD, that there is indeed a weak coupling expansion parameter $1/N$ in
the problem, where $N$ is proportional to $Q_1$ and $Q_5$. $N$ goes to
infinity holding $Ng$ fixed where $g$ is the closed string coupling
constant. The Newton's constant $G$ scales as $1/N^2$.  There is a
partial analogy with large-N baryons of QCD. In section 3, we describe
some relevant aspects of the $U(Q_1)\times U(Q_5)$ supersymmetric
gauge theory that describes the D-brane system corresponding to the
black hole. In section 4 we consider this theory in a phase where, in
the ground state, Wilson lines in the centre of the gauge group as
well as the hypermultiplets condense and then analyze the ground state
structure of the theory. In section 5, we isolate the supersymmetric
gauge invariant degrees of freedom of the low-lying excitation
spectrum at long wave lengths. We find that a combination of the
Wilson lines in the centre of $SU(Q_1)$, (say, for $Q_1>Q_5$) and a
residual invariance under Weyl reflections in $SU(Q_5)$, leads to the
existence of exactly 4 excitations that effectively live on a circle
of radius $RQ_1Q_5$. We perform a quenched average over the different Wilson lines as is standard in the limit of large $N$. The residual Weyl reflections are treated by
using the gauge invariant sewing mechanism
\cite{DVV2,MOTL,BKSEI,DVV1}. This leads to an N=4 supersymmetric
$\sigma$-model in $1+1$ dimensions with central charge $c=6$.  Though
this analysis is performed in the weak coupling limit $g\rightarrow
0$, one may argue, based on the supersymmetry of the model, that the
results are not changed in any drastic way in the black hole regime
where $gN > 1$. We also discuss the failure of the ``unitary gauge
fixing'' and the appearance of $Z_2$ vortices. The $Z_2$ vortex
corresponds to a marginal operator in the effective superconformal
field theory.  Section 6 contains our conclusions and some comments
about the possibility of obtaining a description of black holes in the
M(atrix) theory \cite{BFSS} which seems poised to unify most of the
disparate facts about string theory.

\section{Finite Horizon Area and the `Large-$N$' Limit}
This section has much overlap with some parts of \cite{MALDASTROM,
POLSTROMD}, however, given the importance of the observations, it is
well worth discussing them with a slightly different emphasis. The
black hole solution of IIB string theory (see \cite{MALDATHESIS} for a
review) compactified on a 5-torus, is characterized by 3 integral
chrages $Q_1$, $Q_5$ and $n$. This solution saturates the BPS
bound. $Q_1$ and $Q_5$ are the electric and magnetic type charges
associated with the RR 3-form field strength and $n$ is related to the
momentum $P$ along the 5th circle of radius $R$ which is quantized as
$P=n/R$. The entropy of the black hole turns out to be
\beq 
S = 2\pi \sqrt{Q_1 Q_5 n}\,,
\label{S}
\eeq 
and is related to the horizon area $A$ by $S= A/4\pi G$. Equating
these two expressions, we get an expression for the area of the horizon
of the black hole , $A\sim g^2\sqrt{Q_1Q_5n}$. We have used the fact
that, in units of the string length, Newton's coupling is $G\sim g^2$,
where $g$ is the coupling constant of closed string theory. Now since
the black hole solution has a quantum relevance only if $g\rightarrow
0$, we see that for fixed values of the charges the horizon shrinks to
zero, and the black hole is not macroscopic. This objection can be
averted if we send the charges to infinity in a specific way. Since the
D-brane assembly is described by a $U(Q_1)\times U(Q_5)$ gauge theory,
it is consistent to hold $gQ_1$ and $gQ_5$ fixed, because the gauge
theory in that case has a systematic expansion in powers of $1/
Q_1$ and $1/Q_5$. Hence, a finite horizon area requires holding
$g^2n$ fixed. Further, since the horizon sets the length scale below
which there is no black hole, the relevant regime for the gauge
theory is $(gQ_1gQ_5g^2n)^{1/6} >> 1$. Since $Q_1$ and $Q_5$ both
scale as $1/g$, they are comparable. It is important to realise
that all the 3 charges have to be scaled to infinity in order to
describe a finite area of the horizon. Holding any charge at a finite
value leads to a zero horizon area in the limit of weak closed string
coupling. To facilitate further discussion we introduce the natural
notation $N\sim {Q_1}\sim {Q_5}$.  Needless to say, solving the gauge
theory in the regime $gN>1$ is a difficult problem, however, it is
important to realise that there exits a systematic expansion in the
small expansion parameter $1/N$. For various aspects of the large
$N$ limit of gauge theories and matrix models which are relevant to us
in this section and later on, we refer the reader to \cite{BREZIN}.

Let us indicate the $N$ scaling of some relevant quantities. Newton's
coupling scales as $G\sim 1/N^2$. The mass of the black hole scales as
$M\sim N^2$ and its entropy scales as $S\sim N^2$. This is what we
expect in the semi-classical limit as ${{1/N}{\rightarrow 0}}$.  There
is an instructive analogy with $SU(N)$ QCD in the large $N$ limit. As
is well known the low energy effective lagrangian of QCD is the chiral
model of mesons whose expansion parameter is given by the inverse of
the pion coupling constant ${f_\pi}\sim 1/N$. The baryon is an $N$ quark
bound state interacting via gluons and it is a soliton solution of the
chiral Lagrangian with $M\sim N$. The baryon is analogous to the black
hole that is composed of $N^2$ open string degrees of freedom, which,
as we will subsequently see, arise from the hypermultiplet of $N^2$
strings in the fundamental representation of $U(Q_1)\times
U(Q_5)$. Just like in QCD, where meson-baryon couplings are of order
$(1/N)^0$ and of the same order as the pion kinetic energy term, the
closed string- black hole couplings are also of order $(1/N^2)^0$, of
the same order as the graviton kinetic energy term. In both cases the
interaction is of order one and that is why there is a non-trivial
scattering. Also, the size of the baryon is independent of $N$ and so
is the area of the horizon of the black hole. However the analogy is
partial because the lowest lying collective modes of the baryon are
described by the collective coordinates of the flavour group, and
hence the degeneracy of the ground state does not increase
exponentially. An exponential increase in the number of states is of
course the main feature of the modern basis of black hole
thermodynamics \cite{SUSS,SEN}, at least for the black hole solutions
of string theory. The other significant difference is that the mass of
a D-brane is of order $N$, and hence it can have strong order $N$
interactions with the black hole \cite{STEVE}.

\section{The Supersymmetric Gauge Theory for the D-brane System}
The five dimensional black hole in type IIB string theory compactified
on $T^5$ and carrying charges $Q_1$ and $Q_5$, is modelled by a system
of $Q_1$ D1-branes and $Q_5$ D5-branes \cite{CALMALDA}. This system
is, in turn, described by a $U(Q_1)\times U(Q_5)$ gauge theory in two
dimensions which is the dimensional reduction of an N=1 supersymmetric
$U(Q_1)\times U(Q_5)$ gauge theory in six dimensions, or equivalently,
that of an N=2 supersymmetric theory in four dimensions. In this
section, we briefly describe the massless spectrum of this D-brane
system and write down the relevent terms of the corresponding
2-dimensional gauge theory Lagrangian.

Consider a system of $Q_5$ D5-branes paralle to $x^1,x^2,x^3,x^4,x^5$
and $Q_1$ D1-branes parallel to $x^1$. The excitations of this system
correspond to open strings with the two ends attached to the branes
and there are four classes of such strings: the (1,1), (5,5) (1,5) and
(5,1) strings. In the absence of D1-branes, the part of the spectrum
corresponding to (5,5) strings is the dimensional reduction, to $5+1$
dimensions, of the N=1 $U(Q_5)$ gauge theory in $9+1$ dimensions. In
six dimensions, this consists of a vector multiplet and a
hypermultiplet in the adjoint representation of the gauge group. When
we introduce D1-branes, say along $x^1$, then this gauge theory has to
be further restricted to two dimensions, {\it i.e.}, to the $x^1-t$
space. Now, we also have (1,1) strings and the states coming from
these correspond to the dimensional reduction, to two dimensions, of
N=1 $U(Q_1)$ gauge theory in ten dimensions. The field content
obtained so far is the same as that of N=2 $U(Q_1)\times U(Q_5)$ gauge
theory in 6 dimensions, reduced to 2 dimensions.  However, there are
also other fields comming from the (1,5) and (5,1) strings. These
strings have ND boundary conditions along the coordinates
$x^2,x^3,x^4,x^5$, that are transverse to the D1-branes, but parallel
to the D5-branes. The excitations coming from these strings fall in a
hypermultiplet and their inclusion reduces the N=2 supersymmetry in 6
dimensions to N=1. Since these strings have their ends fixed on
different types of D-branes, the corresponding fields transform in the
fundamental representation of both $U(Q_1)$ and $U(Q_5)$.

A theory describing these excitations at low energies can be easily
written down by first constructing an $N=2$ $D=4$ gauge theory with
gauge group $U(Q_1)\times U(Q_5)$ containing a pair of vector
multiplets and a pair of hypermultiplets in the adjoint
representations of the gauge groups, along with a hypermultiplet in
the fundamental representation of each factor of the gauge group (see,
for example, \cite{West,REV}). The dimensional reduction of this
theory to 2 dimensions then gives the gauge theory describing the
low-energy dynamics of the D-brane system. In the following we will
only write down the few relevant terms of this lagrangian which are
needed for the analysis of the next section. But first, some notation:
The fundamental representation indices are denoted by $a,b,\dots$ for
$U(Q_5)$, and $a',b',\dots$ for $U(Q_1)$. The indices $i,j$ label the
fundamental doublet of $SU(2)_R$ and its generators are denoted by
$\tau^I/2$. The scalar components of the hypermultiplet in the
fundamental representation of the gauge groups are denoted by
$\chi_{ia'a}$ and its spinor components are denoted by $\psi_{a'a}$
and $\t\psi_{a'a}$. For the gauge fields of $U(Q_1)$ and $U(Q_5)$ we
use the notations $A^{(1)s'}_\al$ and $A^{(5)s}_\al$ respectively,
where $\al = 0,1$ and $s,s'$ label the adjoint representations. The
scalar components of the adjoint hypermultiplets are denoted by
$N_i^{(1)}$ and $N_i^{(5)}$, and their fermionic superpartners are
denoted by $\Sigma^{(1)}, \widetilde\Sigma^{(1)}$ and $\Sigma^{(5)},
\widetilde\Sigma^{(5)}$, respectively. Under a gauge transformation,
$\chi_i$ transform as $\chi_i\rightarrow U_1\chi_i U^{-1}_5$ where,
$U_5\in U(Q_5)$ and $U_1\in U(Q_1)$. The relevant terms in the
Lagrangian are the ones involving the hypermultiplets and are given by
\bea
L&=&\int d^2x \Big[
-(D_\al\chi^i)^\d_{aa'}(D^\al\chi_i)_{a'a} 
-\frac{1}{2}\b\psi^{aa'}\g^\al (D_\al \psi)_{a'a}
-\frac{1}{2}{\b{\t\psi}}^{aa'}\g^\al (D_\al \t\psi)_{a'a}
\nonumber\\
&&\quad\qquad 
-{\rm Tr}\left(D_\al N^{i(1)\d}D^\al N_i^{(1)} + 
\frac{1}{2}{\b\Sigma^{(1)}}\gamma^\al D_\al \Sigma^{(1)} +
\frac{1}{2}{\b{\t\Sigma}}^{(1)}\gamma^\al D_\al \t\Sigma^{(1)}\right)
\nonumber\\
&&\quad\qquad
-{\rm Tr}\left(D_\al N^{i(5)\d}D^\al N_i^{(5)} + 
\frac{1}{2}\b\Sigma^{(5)}\gamma^\al D_\al \Sigma^{(5)} +
\frac{1}{2}{\b{\t\Sigma}}^{(5)}\gamma^\al D_\al\t\Sigma^{(5)} \right)
\nonumber\\
&&\quad\qquad
-8g^2D^{(1)s'}_ID^{(1)s'}_I-8g^2D^{(5)s}_ID^{(5)s}_I+ \dots\,\Big]\,,
\label{B-I}
\eea
where the covariant derivatives are 
\beq
\ba{rcl}
(D_\al\chi_i)_{a'a}&=&\del_\al\chi_{ia'a} - iA^{(1)b'}_{\al a'}
\chi_{ib'a} + i \chi_{ia'b} A_{\al a}^{(5)b}\,,
\\{}\\
D_\al N_i^{(1,5)}&=&\del_\al N_i^{(1,5)} + i [N_i^{(1,5)}\,,\,
A_\al^{(1,5)} ]\,, \ea
\label{B-II}
\eeq
and the D-terms are given by
\bea
D^{(1)s'}_I&=&\chi^*_{iaa'}T^{s'}_{a'b'}\tau_{Ij}^i  
\chi^j_{b'a} - \tau_{Ij}^i T^{s'}_{a'b'} [N^{(1)\d}_i\,,\,N^{(1)j}]_{b'a'}
\nonumber\\
&\equiv&{\rm Tr}\left\{T^{s'}(\chi^j \tau_{Ij}^i \chi^\d_i
- \tau^i_{Ij} \,[N^{(1)\d}_i\,,\, N^{(1)j}])\right\}\,,
\label{B-III} \\
D^{(5)s}_I&=&\chi_{ia'a} T^{s}_{ab}\tau_{Ij}^i 
\chi^{*j}_{ba'} -\tau_{Ij}^i T^{s}_{ab} [N^{(5)\d}_i\,,\,N^{(5)j}]_{ba}
\nonumber\\
&\equiv&{\rm Tr}\left\{T^{s}(\chi^{\d j} \tau^i_{Ij}\chi_i
-\tau^i_{Ij} \,[N^{(5)\d}_i\,,\, N^{(5)j}])\right\}\,.
\label{B-IV}
\eea
In the above, we have chosen the signature of the 2-dimensional 
metric as $\{-,+\}$ and the $\gamma$ matrices are given by
$\gamma^0=i\sigma_1$, $\gamma^1=-\sigma_2$, and $\gamma_5=
\gamma^0\gamma^1=\sigma_3$. The conjugate of a spinor is defined by
$\bar\psi=-i\psi^\d\g^0$. The remaining terms of the Lagrangian that
we have omitted, correspond to two vector multiplets and their
couplings to the hypermultiplets. As we will argue, these omitted
fields are not needed for our analysis in the next section.

The gauge theory we have described above corresponds to a system of
D1-branes and D5-branes. However, under T-duality transformations
along directions transverse to both the 1-branes and the 5-branes,
this system is equivalent to a collection of $Q_1$ D5-branes inside
$Q_5$ D9-branes which is described by a gauge theory in (5+1)
dimensions. Our (1+1)-dimensionsal gauge theory is related to this
(5+1)-dimensional theory by dimensional reduction. On the other hand,
under a T-duality along the $x^1$ direction, which is inside both the
D1-branes and the D5-branes, this system is equivalent to a collection
of $Q_1$ D0-branes and $Q_5$ D4-branes. The corresponding gauge theory
is obtained from the theory we will analyize here by dimensional
reduction to (0+1) dimensions. This later theory is relevant in
connection with the matrix theory which contains 0- and 1-branes.
In all these cases the structure of the D-terms remain unchanged.

\section{Analysis of the Gauge Theory Vacuum}
In this section we analyze the classical vacuum of the 2-dimensional
$U(Q_1)\times U(Q_5)$ gauge theory on a circle of radius $R$, in the
limit $g\rightarrow 0$, where $g$ now denotes the gauge coupling
constant.  Our intention, in this section and the next, is to see if
this theory has a phase in which the low-lying excitations admit an
effective description in terms of a conformally invariant
$\sigma$-model on a 4-dimensional moduli space of vacua (and hence
with central charge c=6) living on a circle of radius $Q_1Q_5R$.

First, we select a branch of the vacuum moduli space on which all
fermions as well as the scalars in the vector multiplets are set to
zero but the scalar components of the hypermultiplets, that is,
$\chi_i$, $N^{(1)}_i$ and $N^{(5)}_i$ are non-zero.  Furthermore, we
work in a gauge $A_0=0$. We can also have Wilson lines in the two
gauge groups condensing in the vacuum. These will in general break the
gauge symmetry, except when the Wilson lines are elements of the
centers $Z_{Q_1}$ and $Z_{Q_5}$ of the $SU(Q_1)\times SU(Q_5)$
subgroups of the guage group. We do not consider Wilson lines in the
overall U(1) factors of the gauge groups as this will give a mass to
the fundamental representation fields. For definitness, let us
consider the case $Q_1 >Q_5$. Then it turns out, as we will argue
below, that only Wilson lines in $SU(Q_1)$ are important. Therefore,
we only consider a Wilson lines in the center $Z_{Q_1}$ of $SU(Q_1)$,
which, in a simple parametrization, can be written as $w(px,0)$,
where $p=1,2,..,{Q_1-1}$, and 
\beq 
w(x,0)=e^{i\int_0^x A^{(1)0}_x dx} = 
\left(\ba{cccc}
e^{ix/Q_1R} & 0 & \cdots & 0 \\ 
0 & e^{ix/Q_1R} & & 0 \\ 
\vdots & & \ddots & \vdots \\ 
0 & 0 & \cdots & e^{-i(Q_1-1)x/Q_1R} 
\ea\right)\,.
\label{C}
\eeq

The vacuum is determined by setting the covariant derivatives
(\ref{B-II}) in the presence of the above Wilson lines, as well as the
D-terms (\ref{B-III}) and (\ref{B-IV}) equal to zero. This leads to
the solution
\beq
\chi_{ip}^0(x) = w(px,0) q_i^0(0)\,,
\quad N^{(1)0}_{ip}(x)= w(px,0)n^{(1)0}_i(0) w(0,xp)\,,
\quad N^{(5)0}_i(x)=n^{(5)0}_i(0)\,,
\label{C-I}
\eeq
where, $w(2\pi R,0)\in Z_{Q_1}$ and $q_i^0(0)$, $n^{(1,5)0}_i(0)$
satify the D-term vanishing conditions. This solution is consitent
with supersymmetry transformations and leads to zero variation for the
fermionic fields.  Note that, to write the above solution, we have
picked the point $x=0$ as the reference point and $\chi_{ip}^0(x)$ is the
parallel transport of $q_i^0(0)$ in the presence of the Wilson line $w(px,0)$.

We will now analyze the D-term vanishing conditions: The D-terms in
(\ref{B-III}) and (\ref{B-IV}) are of the form $D^s_I=Tr(T^s A_I)$,
where, $A_I$ is a hermitian matrix and the generators $T^s$ also
include the identity matrix. As a result, $D^s_I=0$ implies $A_I=0$.
Therefore, the vanishing of the D-terms leads to
\bea
&(\chi^j \tau_{Ij}^i \chi^\d_i
- \tau^i_{Ij} \,[N^{(1)\d}_i\,,\, N^{(1)j}])_{a'b'}=0&\,,
\label{C-IIa}
\\
&(\chi^{\d j} \tau^i_{Ij}\chi_i
-\tau^i_{Ij} \,[N^{(5)\d}_i\,,\, N^{(5)j}])_{ab}=0&\,.
\label{C-IIb}
\eea
We want to obtain an explicit solution for the $\chi_i$ by analyzing
these equations for $I=3,2,1$. First, let us make an ansatz for the
adjoint hypermultiplets as
\beq
N^{(1)}_1=N^{(1)}_2\,,\qquad  N^{(5)}_1=N^{(5)}_2\,.
\label{C-III}
\eeq
With this ansatz, the $I=3$ components of (\ref{C-IIa}) and
(\ref{C-IIb}) reduce to
\bea
\chi_1 \chi_1^\d - \chi_2 \chi_2^\d =0
\label{C-IIIa}\\
\chi^\d_1 \chi_1 - \chi^\d_2 \chi_2 =0
\label{C-IIIb}
\eea  
The $Q_1$ dimensional matrix $\chi_1\chi_1^\d$ can be diagonalized
using the gauge group $U(Q_1)$. Since this matrix has rank $Q_5<Q_1$,
it can always be put in the form $diag[v^2_1,\cdots,
v^2_{Q_5},0,\cdots,0]$ with real $v$'s. This breaks the gauge group
$U(Q_1)$ down to $SU(Q_1 - Q_5)$. Note that the Weyl group of $U(Q_1)$
has a broken subgroup $S(Q_5)$ which permutes the non-zero eigenvalues
of $\chi_1\chi_1^\d$, keeping its diagonal form unchanged. As we shall
see, this has an important implication for the theory. Furthermore,
equation (\ref{C-IIIa}) implies that $\chi_2\chi_2^\d$ has the same
form as $\chi_1\chi_1^\d$.

Let us denote the matrix $diag[v_1,\cdots,v_{Q_5}]$ by $V$ and the
$Q_5\times Q_5$ dimensional non-zero blocks of $\chi_i$ by $X_i$.
Now, any generic $X_i$ can be decomposed as $X_i = H_iU_i$, where
$H_i$ is a hermitian matrix and $U_i$ is a unitary matrix. Then, the
discussion so far implies that $H_1=V$ and $H_2=V'$ where, $V'$ is the
same as $V$ modulo arbitrary negative signs for the diagonal elements.
As we shall see later, the existence of this relative negative signs
is crucial for the existance of the solution. 
Furthermore, equation (\ref{C-IIIb}) reduces to $U_1^\d
V^2 U_1 = U_2^\d V^2 U_2$. The matrix $U_2$ can be set to identity by
a $U(Q_5)$ gauge transformation. This fixes the $U(Q_5)$ gauge
completely, because $U_2$ is an arbitrary unitary matrix. This also
means that the moduli we obtain are $U(Q_5)$ invariant. The vanishing
of (\ref{C-IIIb}) then determines $U_1$ to be of the form
$U_1^0=diag[e^{i\theta_1},\cdots e^{i\theta_{Q_5}}]$, and we get 
$\chi_1=VU_1^0$, $\chi_2=V'$.

The $I=2$ components of (\ref{C-IIa}) and (\ref{C-IIb}) reduce to
\beq
\chi_2\chi_1^\d - \chi_1\chi_2^\d =0\,,\qquad
\chi_2^\d\chi_1 - \chi_1^\d\chi_2=0\,,\qquad
\eeq
This implies that $U^0_1=1$, so that in terms $q^0_i$ of (\ref{C-I}),
the vacuum can be parametrized by 
\beq
q^0_1(0) = \left( \ba{cccc}
v_1&      0        & \cdots &  0  \\
       0        &v_2& \cdots &  0  \\
\vdots          &               & \ddots &\vdots\\
   0            &      0        & \cdots & v_{Q_5}\\
   0            &      0        & \cdots &  0    \\
  \vdots        &               &        & \vdots\\   
   0            &      0        & \cdots &  0    
\ea \right)\,,    
\qquad
q^0_2(0) = \left( \ba{cccc}
v'_1    &  0  & \cdots &  0  \\
0      & v'_2 & \cdots &  0  \\
\vdots &     & \ddots &\vdots\\
   0   & 0   & \cdots &v'_{Q_5}\\
   0   & 0   & \cdots &  0    \\
\vdots &     &        & \vdots\\   
   0   & 0   & \cdots &  0    
\ea \right)    
\label{C-IV}
\eeq
where $v_a$ and $v'_a$ can differ by a negative sign. 
The $I=1$ components of (\ref{C-IIa}) and (\ref{C-IIb}) give
\beq
\chi_2\chi_1^\d + \chi_1\chi_2^\d=2[N^{(1)\d}_1\,,\,N^{(1)}_1]\,,\quad
\chi_2^\d\chi_1 +\chi_1^\d\chi_2=2[N^{(5)\d}_1\,,\,N^{(5)}_1]\,,\quad
\eeq
These constrain $N^{(1,5)}_i$ in terms of $\chi_i$. Note that the right
hand side of the above equation is traceless. This in turn constrains 
the $q^0_i(0)$ in (\ref{C-IV}) so that
\beq
\sum_{a=1}^{Q_5} v_a v'_a=0
\eeq
Thus, the vacuum is parametrized by (\ref{C-IV}) subject to the above
constriant. As we will see in the next section, the precise structure
of $N^{(1,5)}_i$ is not relevent to our problem. 

Our starting theory also had a global $SU(2)_R$ symmetry which is
broken by the above solution. As a result, the moduli space also has
three more parameters coming from the $SU(2)_R$ rotations of the doublet
$q^0_{ia}$, where $a=1,\cdots, Q_5 $ labels the diagonal elements in 
(\ref{C-IV})). These give rise to Goldstone bosons in the effective
theory. If we denote this $SU(2)_R$ rotated doublet by $q^0_{a}(0)$,
then in the standard parametization of $SU(2)$, we can write
\beq
q^0_{a}(0)=\left(\ba{c}q^0_{1a}(0)\\q^0_{2a}(0)\ea\right)=
\left(\ba{cc}a&b\\-b^*&a^*\ea \right)\left(\ba{c}v\\v'\ea\right)
\eeq
where, $|a|^2+|b|^2=1$. Below, $q^0_{a}(0)$ always refers to the above
rotated doublet.  

The parametrization (\ref{C-IV}) does not fix the action
of a Weyl group diagonally embedded in the two gauge groups: The Weyl
group of SU(N) acts as a permutation group on the entries of a
diagonal matrix transforming in the adjoint representation of the
group. Therefore, a diagonal combination of the Weyl group $S(Q_5)$ of
$U(Q_5)$ acting from the right, and the subgroup $S(Q_5)$ of the Weyl
group $S(Q_1)$ of $U(Q_1)$ acting from the left will preserve the
diagonal form of $q^0$, while permuting its eigenvalues. Therefore,
the actual moduli space is parametrized by (\ref{C-IV}) with proper
identifications under the action of this diagonal Weyl group. We will
discuss this in more detail in the next section.

\section{The Effective String Theory}
Having parametrized the vacuum, we now study the effective theory for
the low-lying excitations on the moduli space of vacua. This has to be
done carefully because of the presence of the Wilson line in the
vacuum  which affects the periodicity of the fields on $S^1$. For the
fields $\chi_i$, the excitations around the ground state can be
written as  
\beq
\chi_i(x) = w(px,0) q_{ip}(x)\,, \qquad{\rm where},\quad q_{ip}(x) = w(0,px)
q^0_i(x) 
\label{C-V}
\eeq
These are still flat directions for the D-terms. Since the ground
state was defined with respect to a reference point $x=0$, the Wilson
line $w(0,px)$ in $q_{ip}(x)$ is needed to transport back $q^0_i(x)$ to
this reference point. The $q_{ip}(x)$ now also contains the Goldstone
modes coming from the $SU(2)_R$ rotations. Other components of
$\chi_{iaa'}$ do not enter the low-energy description since they are
either gauge degrees of freedom that have been gauged away (we fix the
same ``unitary type'' local gauge on the fluctuations at the point $x$
as we did in the previous section at the point $x=0$), or they do not
correspond to flat directions and hence are massive. The field
$\chi(x)$ appears in the original theory and is periodic on a circle
of radius $R$. The decomposition in (\ref{C-V}) implies that $q^0(x)$
is also periodic on the same circle while $q_p(x)$ is periodic on a
circle of radius $RQ_1$. Since the gauge is fixed with reference to
the point $x=0$, $q_{ip}(x)$ is gauge invariant under a local gauge
transformation and hence it is the true physical degree of
freedom. Substituting (\ref{C-V}) in the kinetic energy term in
(\ref{B-I}), one can easily see that while $q^0_i(x)$ has a mass term
coming from the Wilson line, $q_{ip}(x)$ is massless.

We now have the situation of a field defined with twisted 
boundary conditions on a circle of radius $R$. In the limit of large $Q_1$
the gauge field average can be replaced by a quenched average over the 
twisted boundary conditions which is familiar from the Eguchi-Kawai reduction at large $N$ \cite{BREZIN}. This fact leads to a single  
massless field $q_i(x)$ which is periodic on the circle of radius $Q_1R$.  
Hence we have an
effective theory of $Q_5 +3 $ massless fields on a circle of radius
$RQ_1$. The action (\ref{B-I}) also shows that all these fields are
accompanied by their fermionic superpartners. A similar discussion
applies to the field $N^{(1)}$. However, since $N^{(1)}$ is in the
adjoint representation and the Wilson line is in the center of the
group, these excitations live on circle of radius $R$. Therefore,
their momenta are quantized in units of $1/R$ (as opposed to $1/RQ_1$
for the fundamental representation hypermultiplet) and hence, at low
energies, they decouple for the theory. The same is the case with all
the other fields we have not considered above.

Let us now consider some further periodicity properties of this
solution. Since $q_i(x)$ are defined modulo the action of the
diagonal Weyl group $S(Q_5)$, we have
\beq
q^0_i(x+2\pi R) = S q_i^0(x) S^\d\,\,,\qquad S\in S(Q_5)
\label{C-VI}
\eeq
If we label the diagonal elements of $q_i^0(x)$ by the index
$a=1,\cdots, Q_5$, then this means $q_{ia}^0(x+2\pi R) =q_{ib}^0(x)$.
This property can be easily taken into account if we sew
\cite{DVV2,MOTL} the functions $q_{ia}^0(x)$ into a single function
$f^0_i(x)$ with period $2\pi RQ_5$, defined over a circle of radius
$RQ_5$. In fact, $q_{ia}^0(x)=f^0_i(x+{2\pi aR})$ so that for any
gauge invariant function $F\left(q^0_{ia}(x)\right)$ of the moduli, we
have
\beq
\sum_{a=1}^{Q_5}\int_0^{2\pi R}dx\, F\left(q^0_{ia}(x)\right)
= \int_0^{2\pi R Q_5}dx \,F\left(f^0_i(x)\right)
\label{C-VII}
\eeq
Now, instead of the functions $q_{ia}^0(x)$, we can sew the $Q_5$
functions $q_{ia}(x)$ in (\ref{C-V}) into a single function
$f_i(x)$. Taking into account the three $SU(2)_R$ Goldstones, $f_i(x)$
describes precisely 4 massless modes on a circle of radius $RQ_1Q_5$.

Upto now we have ignored the presence of the Wilson lines in the
centre of $SU(Q_5)$. In the previous section we have already indicated
that the gauge fixed moduli are invariant under $U(Q_5)$
transformations. Hence they simply do not see these Wilson lines. It
is important to realize that this is true only within the moduli space
approximation one is working with.

In the above discussion we have only included hypermultiplets in the
fundamental representation since it is for these fields that the
radius of the circle $S^1$ is dailated and momentum is quantized in
units of $1/RQ_1Q_5$. In the limit of large $Q_1$ and $Q_5$, all other
fields are much heavier than these and decouple in the long wavelength
limit. Therefore they do not contribute to the spectrum of the
low-lying excitations of the theory. Now that we have 4 massless
bosons and their superpartners with a flat metric on the moduli space,
we have an $N=4$ SCFT with central charge $c=6$. The black hole in its
ground state is decribed by states at level $Q_1Q_5n$ in this
conformal field theory. This leads to the correct entropy formula
(\ref{S}) for the black hole \cite{MALDASUSS}. Excited states of the
black hole correspond to the presence of both left and right moving
oscillations of the effective string.

Our analysis has been performed in the weak coupling limit of
the gauge theory while the black hole corresponds to the strong
coupling limit. However, we note that 2-dimensional gauge theory is
ultraviolet finite upto normal ordering of the Hamiltonian. Moreover,
the moduli space we obtain is a very special hyperKahler manifold
(related to N=4 supersymmetry of the theory) in 4 dimensions: the flat
space. If there are corrections to this metric, they will become more
and more important as the coupling grows. However, the hyperKahler
geometry is very restrictive \cite{ALFR} and it appears that there are
no such corrections to interpolate between the flat space and some
other non-trivial hyperKahler geometry. Thus we can conclude that our 
moduli space survives in the strong coupling limit.

We now discuss issues related to the Weyl symmetry in some detail. In
the above discussion we considered elements of $S(Q_5)$ with the
longest cycle. These lead to the largest dilation of the radius:
$RQ_1\rightarrow RQ_1Q_5$. For large $Q_5$, these configurations are
more favourable since their entropy grows much faster with $Q_5$ than
the entropy of configurations corresponding to smaller cycles of the
permutation group. For example, compare the entropy for the largest
cycle of length $Q_5$ with that of two cycles of length $Q_5/2$. The
entropy of the largest cycle is greater than that of the shorter
cycles, clearly indicating that when one
averages over all cycles of the Weyl group, the leading contribution
comes from the largest cycle. We also note that the relative
contribution of a cycle of length $Q_5$ and cycles of length $Q_5-1$
and $1$ is suppressed by a factor of $1/Q_5$. This situation is
relevant to the case when the cycle of length $1$ shrinks to zero,
{\it i.e.}, when two adjacent eigenvalues cross each other.

Let us now discuss the important issue of coincident eigenvalues. As
is well known the Weyl measure in the functional integral over the
moduli space \cite{WEYL} vanishes when two eigenvalues coincide.
Coincident eigenvalues are a subleading $1/N$ effect and hence their
effects are proportional to the string coupling $g$. When the
eigenvalues coincide, the unitary gauge condition on the fluctuations
cannot be fixed and it signals the appearance of topological objects
in the space-time under consideration \cite{THOOFT}. In our problem,
since the spacetime is 2-dimensional (here, we assume an analytic
continuation to euclidean space), the topological object at a point on
this surface where the eigenvalues coincide is a vortex with a string
singularity. It is like a vortex in a $SO(3)$ theory and the vortex
charges are given by the elements of the homotopy group $\pi
_1(SU(2)/Z_2)=Z_2$.  Hence the string singularity is a square root
branch cut in the world sheet emanating from the position of the
vortex. It is natural to introduce a local operator on the world sheet
describing the $Z_2$ vortex. Such a construcation has been described
in \cite{DVV1} using twist operators and their corresponding spin
fields \cite{DFMS}. However, in our case since the SCFT has c=6, we
get a marginal operator \cite{SWtalk} that creates the $Z_2$
vortex. In general, when $n$ eigenvalues coincide, one is naturally
lead to $Z_n$ vortices characterized by
$\pi_1(SU(n)/Z_n)=Z_n$. However, the higher twist operators ($n\geq 
3$) are relevant and hence tachyonic. From supersymmetry, one expects
that such operators are not allowed in the theory and are projected
out. 

The appearance of the marginal operator for the $Z_2$ vortex may
cause some concern. In the strict $N=\infty$ limit, it is clear
that the SCFT has a target space $R^4$ which is not renormalized at
strong coupling. In this case, we expect the calculations in
\cite{DMW,DASMATH} to be valid in the strongly coupled black hole
regime.  However, even for a very large value of $N$, not strictly
equal to infinity, the target space at weak coupling ($g\rightarrow
0$) is the orbifold $R^4/Z_2$ which is a direct consequence of the
existance of this operator. At strong coupling, we expect the full
space to be described by the Eguchi-Hanson metric. 
In light of the excellent agreement obtained in \cite{DASMATH}, it is unlikely that such a SCFT describes the Hilbert space of a single black hole. One resolution is to interpret the marginal operator as an interaction in a second quantized theory of blac
k holes \cite{FGS}.

\section{Discussion and Concluding Remarks}
In this paper we have discussed the derivation of the low energy
effective string picture for the 5-dimensional black hole in the
framework of the effective gauge theory that describes the assembly of
a large number of D1-branes and D5-branes that form the black hole. It
turns out that in the limit of large $N\sim Q_1\sim Q_5$, the
effective string corresponds to the gauge invariant collective modes
of the condensed $(1,5)$ and $(5,1)$ open strings in the fundamental
representation of the gauge group which mediate the interactions of
the branes.  Given the fact that the effective theory is a flat four
dimensional sigma model with N=4 supersymmetry, and that hyperKahler
geometry is very restrictive, we expect the weak coupling answers to
persist beyond weak coupling. Our work is one further step in the
direction of modelling of black holes by D-brane constituents.

It should be mentioned that though the picture presented here
resembles the one suggested in \cite{MALDASUSS}, the two differ in
some essential ways. According to \cite{MALDASUSS}, D-branes are
joined to form a long brane which is multiply wound around $S^1$. This
picture amounts to condensing Wilson lines in the the two gauge groups
such that their eigenvalues are the $Q_1th$ and $Q_5th$ roots of unity
\cite{GPOL}. Such Wilson lines are seen by all fields and therefore,
both the fundamental and the adjoint representation fields live on a
circle of larger radius. However, in our case, the Wilson line is in
the centre of the group and only the collective modes of the
hypermultiplets in the fundamental representaion are described by a
theory on the circle of radius $RQ_1Q_5$. All other fields have higher
momenta and their excitations decouple at sufficiently low energies.

The system of $Q_1$ D1-branes and $Q_5$ D5-branes is equivalent to a
system of $Q_5$ D1-branes and $Q_1$ D5-branes under T-duality in the
internal dimensions, and hence the two types of branes can be treated
symmetrically. 

We would like to mention that our treatment of the the D-flatness 
conditions used an explicit anzatz for the hypermultiplet fields in the 
adjoint representation of the gauge groups. An improved treatment which uses the full set of fields will be presented in a future communication
\cite{FS}.

We conclude with a comment on matrix theory. As was indicated in
section 3, we can use T-duality transformations to transform the
$D1,D5$-brane system to a $D0,D4$-brane system, which is described by
an N=8 SUSY Yang-Mills theory in $0+1$ dimensions based on the gauge
group $U(Q_1)\times U(Q_5)$.  The structure of the hypermultiplet is
of course the same as before.  We know that the $D_0-D_4$ sytem exists
in M(atrix) theory \cite{BKSEISH}, and one would like to ask whether
the matrix model that we have mentioned can be derived as an effective
description from M(atrix) theory. This is an important issue because
any fundamental theory of quantum gravity should explain black holes.

\vspace*{.4cm} 

\noindent{\Large\bf{Acknowledgement}}

\vspace*{.3cm} 

We would like to thank G. Mandal, K.S.Narain and A. Sen for many very useful and
enjoyable discussions. We would also like to thank L. Alvarez-Gaume, M.Blau,
J. Maldacena, L. Susskind, G.Thompson and E. Verlinde for
discussions on some aspects of this work. Finally one of us (SW) would
like to acknowledge the `CERN Duality Workshop', for the opportunity
to lecture on the subject of black holes last winter and G. Veneziano
and A. Schwimmer for some thoughtful questions. 

\vspace*{.4cm} 

\noindent{\Large\bf{Note Added}}

\vspace*{.3cm} 

The appearance of the marginal operator corresponding to the $R^4/Z_2$
orbifold has also been observed in \cite{DVV3}. We would like to thank
T. Banks and P. Horava for pointing this out to us.

\end{document}